\documentstyle[preprint, aps]{revtex}

\draft
\tighten

\begin{document}
\title{Stability of strangelet at finite temperature}
\author{Yun Zhang$^{1,3}$ and Ru-Keng Su$^{2,1}$}
\address{$^1$Department of physics, Fudan University, Shanghai 200433 , P. R. China\\
$^2$China Center of advanced Science and Technology (World Laboratory)\\
P. O. Box 8730, Beijing 100080, P. R. China \\
$^3$Surface Physics Laboratory (National Key Laboratory), Fudan University,\\
Shanghai 200433 , P. R. China }
\maketitle

\begin{abstract}
Using the quark mass density- and temperature dependent model, we have
studied the thermodynamical properties and the stability of strangelet at
finite temperature. The temperature, charge and strangeness dependences on
the stability of strangelet are investigated. We find that the stable
strangelets are only occured in the high strangeness and high negative
charge region.
\end{abstract}


\section{Introduction}

The study of small lumps of strange quark matter, called strangelets, plays
an important role for research the quark gluon plasma (QGP) in recent
relativistic heavy ion collision (RHIC) experiments. The reason is that
although many signatures of QGP such as $J/\Psi $ suppression, strangeness
enhancement, thermal dilepton electro magnetic radiation, etc., have been
found \cite{1,2}, but it is still ambiguous because these signatures can
also be explained by hadron gas \cite{3}. To search an unambiguous signature
of QGP is the key for RHIC experiments. The strangelet, as was argued by
Greiner et. al \cite{4}, is a good candidate which could serve as an
unambiguous signature for the QGP.

The essential problem for detectability of strangelet in RHIC experiments is
to study its stability during the formation of QGP. Employing MIT\ bag
model, many authors discussed this problem \cite{5,6,7,8}. They came to the
same conclusion that the electric charge and the strangeness fraction are of
vital important to the experimental searches of strangelets because these
two elements affect a change of stability remarkably. But the effect of
charge on stability is different for different authors. Jaffe and his
co-workers \cite{5,6,7} argued that the strangelet is slightly positive
charged. They considered strong and weak decay by nucleon and hyperon
emission together and concluded that the stable strangelets will have a low
but positive change to mass ratio. Contrary to Jaffe et. al, after
considering the initial condition of possible strangelet production in RHIC
carefully, Greiner and his co-workers \cite{8} argued that strangelets are
most likely highly negative charged. Here we hope to emphasize that above
discussions are limited in the framework of MIT bag model and at zero
temperature.

Since the quark deconfinement phase transition can occur at high temperature
or/and high density only, it is of interest to extend the investigation for
the stability of strangelets to finite temperature. This is the objective of
this paper.

However, MIT bag model is a permanent quark confinement model because the
confined boundary condition does not change with temperature. In principle,
one can not use this model to study the phase transition of QCD directly.
The best that we can do is to employ a model which can almost reproduce the
properties of strange quark matter obtained by MIT\ bag model, but in which
the quark confinement is not permanent. The quark mass density- and
temperature-dependent model (QMDTD) \cite{9,10} suggested by us is one of
such candidates.

Many years ago, a quark mass density-dependent model (QMDD)\ was suggested
by Fowler, Raha and Weiner \cite{11} and then it was employed by many
authors to discuss the properties of strange quark matter \cite{12,13,14,15}%
. According to the QMDD\ model, the masses of $u,d$ quarks and strange
quarks (and the corresponding anti-quarks) are given by 
\begin{eqnarray}
m_q &=&{\frac B{3n_B}},\hspace{0.8cm}(q=u,d,\bar{u},\bar{d}),  \label{su1} \\
m_{s,\bar{s}} &=&m_{s0}+{\frac B{3n_B}},  \label{su2}
\end{eqnarray}
where $n_B$ is the baryon number density, $m_{s0}$ is the current mass of
the strange quark and $B$ is the vacuum energy density. As was proved by ref.%
\cite{13}, the properties of strange matter in the QMDD\ model are nearly
the same as those obtained in the MIT bag model. In fact, it is not
surprised if one notices that the confinement mechanism of MIT\ bag model is
almost the same as that of QMDD\ model \cite{9}.

But when we employ the QMDD\ model to discuss the properties of strangelet,
a lot of difficulties emerge \cite{9,10}. At first, the radius of strangelet
decreases as the temperature increases; Secondly, it can not mimic the
correct phase diagram of QCD because the temperature $T$ tends to infinite
when $n_B\rightarrow 0$. To overcome these difficulties, we suggest a QMDTD\
model \cite{9,10}. Instead of a constant $B$ in QMDD\ model, we argue that $%
B $ is a function of temperature and introduced an ansatz \cite{10} 
\begin{eqnarray}
B(T) &=&B_0\left[ 1-a\left( \frac T{T_c}\right) +b\left( \frac T{T_c}\right)
^2\right] ,0\leq T\leq T_c,  \label{15} \\
B(T) &=&0,T>T_c,  \label{15-1}
\end{eqnarray}
where $B_0$ is the vacuum energy density inside the bag (bag constant) at
zero temperature, $T_{c\mbox{
}}=170\mbox{MeV}$ is the critical temperature of quark deconfinement phase
transition, and $a$, $b$ are two adjust parameters. Since $B$ is zero when $%
T=T_c$, a condition 
\begin{equation}
1-a+b=0  \label{16}
\end{equation}
is imposed and only one parameter $a$ can be adjusted. As pointed out by ref.%
\cite{10}, in order to satisfy two physical conditions of strangelet at
finite temperature, namely, (1), the radius of the strangelet must increases
when the temperature rises, (2), the energy of the strangelet must increases
when the temperature rises, the parameter $a$ is restricted in a small
range: 
\begin{equation}
0.65\leq a\leq 0.8.  \label{17}
\end{equation}

In this paper, we fix the value of two parameters $a$, $b$ in this suitable
range 
\begin{equation}
a=0.65,b=-0.35.  \label{18}
\end{equation}
With these parameters set, we use the QMDTD\ model to discuss the
thermodynamical properties of strangelet, especially, to investigate the
stability of strangelet via strong hadron emission and weak hadronic decay.
We will study the effects of temperature, charge and strangeness fraction on
the stability and hope that our study can have impact to the detectability
of the strangelet.

The organization of this paper is as follows. In the following section, we
give the formulae of thermodynamical calculations for QMDTD\ model. The
results of thermodynamical properties of strangelet are presented in section
3. In section 4, we will discuss the stability of strangelet via possible
strong and weak decays. The last section is a summary.

\section{Thermodynamical formulae}

To calculate the dynamical and thermodynamical quantities of the strangelet,
we must look for the density of states first. The density of states of a
spherical cavity in which the free particles be contained can be expressed
as 
\begin{equation}
\rho (k)={\frac{dN(k)}{dk}},  \label{a}
\end{equation}
where $N(k)$ is the total number of particle states and can be written in
terms of dimensionless variable $kR$ as 
\begin{equation}
N(k)=A(kR)^3+B(kR)^2+C(kR),  \label{2}
\end{equation}
where $R$ is the radius of the bag. The three terms of the right hand side
of Eq.(\ref{2}) refer to the contributions of the volume, surface and
curvature, respectively. The coefficients $A$, $B$ and $C$ are expected to
be very slow varying functions of $kR$, and their expressions are model
dependent. For MIT bag model, these coefficients have been obtained by
numerical calculations in our previous paper \cite{16}. The volume term $A$
is a constant, and it has the value of 
\begin{equation}
A=\frac{2g}{9\pi },  \label{4}
\end{equation}
where $g$ is the total degeneracies. For example, it is the total number of
spin and color degrees of freedom for a quark with flavor treated
separately. The surface term $B$ is 
\begin{equation}
B\left( \frac mk\right) =\frac g{2\pi }\left\{ \left[ 1+\left( \frac mk%
\right) ^2\right] \tan ^{-1}\left( \frac km\right) -\left( \frac mk\right) -%
\frac \pi 2\right\} ,  \label{5}
\end{equation}
where $m$ is the mass of quark. Equations (\ref{4}) and (\ref{5}) are in
good agreement with the ones given by multi-reflection theory \cite
{16-1,16-2}. The curvature term $C$ can not be evaluated by this theory
except for the two limiting case: $m\rightarrow 0$ and $m\rightarrow \infty $%
. Madsen proposed that \cite{16-3} 
\begin{equation}
\widetilde{C}({\frac mk})={\frac g{2\pi }}\left\{ \frac 13+\left( {\frac km}+%
{\frac mk}\right) \tan ^{-1}{\frac km}-{\frac{\pi k}{2m}}\right\} .
\label{6}
\end{equation}
But as was pointed out by ref.\cite{16}, the beat fit of numerical data for
curvature term is 
\begin{equation}
C({\frac mk})=\widetilde{C}({\frac mk})+\left( {\frac mk}\right) ^{1.45}{%
\frac g{3.42\left( {\frac{\displaystyle m}{\displaystyle k}}-6.5\right)
^2+100}.}  \label{Our-C}
\end{equation}

Now we are in the position to calculate the thermodynamical quantities of
strangelets for QMDTD model. The thermodynamical potential $\Omega $ is 
\begin{equation}
\Omega =\sum_i\Omega _i=-\sum_i\frac{g_iT}{(2\pi )^3}%
\displaystyle \int %
_0^\infty dk{\frac{dN_i}{dk}}\ln \left( 1+e^{-\beta (\varepsilon _i(k)-\mu
_i)}\right) ,  \label{omiga}
\end{equation}
where $i$ stands for $u,d,s$ (or $\bar{u},\bar{d},\bar{s}$ ) quarks, $g_i=6$
for quarks and antiquarks. ${%
{\displaystyle {dN_i \over dk}}%
}$ is the density of states\ for various flavor quarks, it is given by eqs.(%
\ref{a})-(\ref{Our-C}). $\mu _i$ is the corresponding chemical potential
(for antiparticle $\mu _{\bar{i}}=-\mu _i$). 
\begin{equation}
\varepsilon _i(k)=\sqrt{m_i^2+k^2}  \label{ek}
\end{equation}
is the single particle energy and $m_{i\text{ }}$is mass for quarks and
antiquarks.

According to the QMDTD\ model \cite{9,10}, the masses of quarks are 
\begin{eqnarray}
m_{u,\bar{u},d,\bar{d}} &=&{\frac{B_0}{3n_B}}\left[ 1-a\left( \frac T{T_c}%
\right) +b\left( \frac T{T_c}\right) ^2\right] ,0\leq T\leq T_c,  \nonumber
\\
m_{u,\bar{u},d,\bar{d}} &=&0,T\geq T_c,  \label{mq} \\
m_{s,\bar{s}} &=&m_{s0}+{\frac{B_0}{3n_B}}\left[ 1-a\left( \frac T{T_c}%
\right) +b\left( \frac T{T_c}\right) ^2\right] ,0\leq T\leq T_c,  \nonumber
\\
m_{s,\bar{s}} &=&0,T\geq T_c,  \label{ms}
\end{eqnarray}
where $m_{s0}$ is the current mass of the strange quark matter, $n_B$ is the
baryon number density 
\begin{equation}
n_B=A/V,  \label{14}
\end{equation}
and $A$ is the baryon number of the strangelet, $V=%
{\displaystyle {4 \over 3}}%
\pi R^3$ is the volume of the strangelet. Using the standard statistical
treatment, and noticing that $\Omega $ is not only a function of
temperature, volume and chemical potential, but also of density, it can be
proved that the total pressure $p$ and the total energy density $\varepsilon 
$ are given by \cite{13,14,9,10}

\begin{equation}
p=-{\frac 1V}\left. {\frac{\partial (\Omega /n_B)}{\partial (1/n_B)}}\right|
_{T,\mu _i}=-{\frac \Omega V}+{\frac{n_B}V}\left. {\frac{\partial \Omega }{%
\partial n_B}}\right| _{T,\mu _i},  \label{18-1}
\end{equation}

\begin{equation}
\varepsilon ={\frac \Omega V}+\sum_i\mu _in_i-{\frac TV}\left. {\frac{%
\partial \Omega }{\partial T}}\right| _{\mu _i,n_B}.  \label{18-2}
\end{equation}
The number density of each particle can be obtained by means of 
\begin{equation}
n_i=-{\frac 1V}\left. {\frac{\partial \Omega }{\partial \mu _i}}\right|
_{T,n_B}.  \label{19}
\end{equation}
At finite temperature, we must include the contributions of the
anti-particles, therefore, the baryon number for $i$ quark is given by 
\begin{equation}
\Delta N_i=(n_i-n_{\bar{i}})\times V=\frac{g_i}{(2\pi )^3}%
\displaystyle \int %
_0^\infty dk{\frac{dN_i}{dk}}\left( \frac 1{\exp [\beta (\varepsilon _i-\mu
_i)]+1}-\frac 1{\exp [\beta (\varepsilon _i+\mu _i)]+1}\right) .  \label{20}
\end{equation}

The strangeness number $S$ of the strangelet reads 
\begin{equation}
S=\Delta N_s,  \label{21}
\end{equation}
the baryon number $A$ of the strangelet satisfies 
\begin{equation}
A={\frac 13}(\Delta N_u+\Delta N_d+\Delta N_s)\hspace{0in}.  \label{22}
\end{equation}
The electric charge $Z$ of the strangelet is 
\begin{equation}
Z=\frac 23\Delta N_u-%
{\displaystyle {1 \over 3}}%
\Delta N_d\hspace{0in}-%
{\displaystyle {1 \over 3}}%
\Delta N_s.  \label{23}
\end{equation}
At finite temperature, the stability condition of strangelets for the radius
reads

\begin{equation}
\frac{\delta F}{\delta R}=0.  \label{13}
\end{equation}
where the free energy $F$ of strangelet is, 
\begin{equation}
F=E-T\tilde{S},  \label{13-1}
\end{equation}
$E=\varepsilon V$ is the total energy, and 
\begin{equation}
\tilde{S}=\sum_i\tilde{S}_i=-\sum_i\left. {\frac{\partial \Omega }{\partial T%
}}\right| _{\mu _i,n_B}  \label{13-2}
\end{equation}
is the entropy.

Given strangeness number, baryon number and electric charge, for any
strangelet, we can calculate chemical potentials for $u,d,s$ quarks
self-consistently by the equations (\ref{21}),(\ref{22}),(\ref{23}) and (\ref
{13}). Then the thermodynamic potential, free energy and the stable radius
of the strangelet at finite temperature can be obtained self-consistently.

\section{Thermodynamical properties of strangelet}

The numerical calculations have been done with the parameters set:

\begin{equation}
B_0=170\mbox{MeV}\mbox{fm}^{-3},m_{s0}=150\mbox{MeV},T_c=170\mbox{MeV},
\label{25-1}
\end{equation}
and the possible area for strangelet in our calculations is chosen as 
\begin{eqnarray}
S &>&0,  \label{25} \\
Z &\geq &-A,  \label{26} \\
S+Z &\leq &2A.  \label{27}
\end{eqnarray}
We calculate the free energy of the strangelet first. Figure 1 shows the
free energy per baryon $F/A$ in a $S-Z$ plane for all possible strangelets
with $A=5$ and $T=20\mbox{MeV}$. This is a set of downward protruding
curves. The corresponding curves but for $T=50\mbox{MeV}$ are shown in
figure 2. Comparing these two figures, we find that the positions of dots
appeared in figure 2 are lower than the corresponding positions (with the
same strangeness number $S$ and charge $Z$) in figure 1. For example, for $%
S=3$ and $Z=-2$, the free energy per baryon of the dot is $F/A=1064.9%
\mbox{MeV}$ in figure 1, but reads $F/A=898.9\mbox{MeV}$ in figure 2. For
fixed $S$ and $Z$, the free energy of the strangelet decreases when
temperature increases. This result is reasonable because the entropy
increases with the temperature, the term $T\tilde{S}$ in eq.(\ref{13-1})
increases considerably and $F$ decreases. We will see in the next section
this result affects on the decay of the strangelet remarkably.

To illustrate our result transparently, we draw the $F/A$ vs $S$ curves for
fixed $A=5,Z=-5$ but different temperatures $T$ $=20\mbox{MeV}$ and $50%
\mbox{MeV}$ in figure 3, respectively. We find that the minimum of these two
curves are located at the same strangeness $S=4$ but with different free
energies.. The whole curve for $T=20\mbox{MeV}$ is located on the upper
position of the curve for $T=50\mbox{MeV}$. We also draw the $F/A$ vs $S$
curve for $A=10,Z=-10$ and $T=50\mbox{MeV}$ in figure 4. Compared with
figure 3, the shape of the curve in figure 4 almost does not changes except
the position of the minimum is changed to $S=8$. These results are similar
to that one given by \cite{17}.

Now we turn to study the charge and strangeness dependences of the radius of
strangelet. We draw $F/A$ vs $R$ curves with fixed $A=5$ ,$Z=1$ and $T=50%
\mbox{MeV}$ but different $S=2,6$ and $9$ in figure 5, respectively. The
stable radiuses given by Eq.(\ref{13}) for different strangeness are
different. Figure 5 shows that the stable radius changes from $1.64\mbox{fm}$
to $1.615\mbox{fm}$ when strangeness number changes from $2$ to $9$. The
same curves for $A=5$ ,$Z=-4$ and $T=50\mbox{MeV}$ but different $S=2,14$
are shown in figure 6. We see that the stable radius become $1.67\mbox{fm}$
for $S=2$ and $1.62\mbox{fm}$ for $S=14$. Therefore, we come to a conclusion
that the stable radius of strangelet decreases when $S$ increases.

To illustrate the charge dependence of the stable radius, we draw $F/A$ vs $R
$ curves with fixed $A=5$ ,$S=2$ and $T=50\mbox{MeV}$ but different electric
charge $Z=1,5$ and $8$ in figure 7, respectively. The stable radius
increases from $1.64\mbox{fm}$ to $1.69\mbox{fm}$ when the electric charge
rises from $1$ to $8$. The same curves for $A=5$ ,$S=9$ and $T=50\mbox{MeV}$
but different charge $Z=1,-4$ are shown in figure 8. We see that the stable
radius changes from $1.615\mbox{fm}$ to $1.605\mbox{fm}$ when $Z$ decreases
from $1$ to $-4$. We find that the stable radius of strangelet increases
with electric charge .

\section{Stability of the strangelet}

In this section, we follow the line of ref.\cite{8} to investigate the
stability of the strangelet and extend their study to finite temperature by
using the QMDTD\ model.

\subsection{Strong decay and unstable strangelet}

As was pointed out by refs.\cite{21,22}, small clusters of strange matter
are most favoured for detection. As two examples, hereafter we study two
cases with $A=5$ and $A=10$, respectively.

Instead of the binding energy at zero temperature, we calculate the free
energy per baryon of the possible strangelet at finite temperature first.
Figure 9 and 10 show the free energy per baryon of all possible strangelets
as a function of the strangeness $S$ at the temperature $T=50\mbox{MeV}$ but
for $A=5$ and $A=10$, respectively. The solid lines in these two figures
connect the masses of the nucleon, $\Lambda ,\Xi $ and $\Omega $. As a first
cut for potential candidates of stable strangelets, these ones lying above
this line can (or probably will) completely decay to the pure hadron state
via strong processes, and only these ones beneath the line will be possible
for metastable or stable strangelets \cite{8}.

We consider the influence of temperature now. As shown in last section, the
free energy per baryon increases when temperature decreases. The positions
of the dots will raise and many dots will across the line and become
unstable when temperature decreases. Figure 11 shows this result clearly.
The figure 11 is the same as figure 9 except for temperature $T=20\mbox{MeV}$%
. Comparing these two figures, we find many dots across the line when
temperature decreases from $50\mbox{MeV}$ to $20\mbox{MeV}$. At low
temperature, more strangelets can decay to the pure hadronic state by strong
process.

Here we must emphasize that although any strangelet initially formed under
this line can not decay to a pure hadron state, it is still possibly
unstable because it can decay to a hadron and another strangelet changing
baryon number, strangeness number and charge \cite{8}. We will look for
possible strong decays, i.e. single baryon \{n, p, $\Lambda ,\Sigma
^{+},\Sigma ^{-},\Xi ^0,\Xi ^{-}$ and $\Omega $\} emission and mesonic
decays at the finite size configuration at finite temperature.

The baryon number $A$, strangeness number $S$ and electric charge $Z$ are
conserved in the strong process. A general expression of a strong baryon
decay for a strangelet $Q(A,S,Z)$ can be written as 
\begin{equation}
Q(A,S,Z)\rightarrow Q(A-1,S-S_x,Z-Z_x)+x(1,S_x,Z_x).  \label{28}
\end{equation}
And this process is allowed if the free energy balance of the corresponding
reaction is 
\begin{equation}
F(A,S,Z)>F(A-1,S-S_x,Z-Z_x)+m_{x,}  \label{29}
\end{equation}
where $F$ stands for the total free energy of the strangelet and $x$ stands
for a baryon with strangeness number $S_x$ and electric charge $Z_x$. Our
results of numerical calculation are shown in figures 12, 13 and 14. For $A=5
$ and $T=50\mbox{MeV},$ all strangelets including the ones lying above the
lines of figure 9 and the ones which satisfy the inequality (\ref{29}) are
drawn by the circles in figure 12. These strangelets can strong decay and
are unstable. The unstable strangelets for $A=10$ and $T=50\mbox{MeV}$ are
drawn by circle in figure 13. Figure 12 and figure 13 show that the
strangelets with small strangeness number $S$ and situated in the left side
of the figure are unstable.

To study the temperature effect, the same figure for $A=5$ and $T=20%
\mbox{MeV}$ are shown in figure 14. Comparing figure 12 and figure 14, we
find that many strangelets become unstable when temperature decreases. Lower
temperature is in favour of the strong decay of the strangelet.

\subsection{Weak decay and metastable strangelet}

According to the definition of Greiner et. al \cite{8}, ''a strangelet is
called metastable in the following if its energy lies under the correspond
(free) hadronic matter of the same baryon number, charge and strangeness,
and if it can not emit a single hadron or multiple hadrons by strong
processes.'' At finite temperature, instead of energy, we use free energy. A
metastable strangelet can then only decay via weak process like the
nonleptonic (hadronic) decays. In this sub-section, we study all possible
weak hadronic decay for metastable strangelets.

In the weak process, the baryon number $A$ and the electric charge $Z$ are
conserved, but the strangeness number $S$ is not conserved. For the weak
decay, $\Delta S=\pm 1$. Therefore, a general expression of a weak baryon
decay for a strangelet $Q(A,S,Z)$ can be written as 
\begin{equation}
Q(A,S,Z)\rightarrow Q(A-1,S-S_x-1,Z-Z_x)+x(1,S_x,Z_x).  \label{30}
\end{equation}
This process is allowed if the free energy balance of the corresponding
reaction is 
\begin{equation}
F(A,S,Z)>F(A-1,S-S_x-1,Z-Z_x)+m_{x.}  \label{31}
\end{equation}

The metastable strangelets which satisfy the inequality (\ref{31}) are shown
in figure 12, 13 and 14 with filled circle. We find that they are almost
situated in the area with negative charge (or slightly positive charge) and
high strangeness number. In particular, high strangeness is in favour of the
stability of strangelet. As shown in figures 12, 13 and 14, a minimal
strangeness $S_c$ above which ($S>S_c$) strangelets are metastable or stable
exists. The values of the minimal strangeness $S_c$ are $5$ in figure 12 ($%
A=5,T=50\mbox{MeV}$) and $10$ in figure 13 ($A=10,T=50\mbox{MeV}$). Defining
strangeness fraction $f_{s\text{ }}$ as 
\begin{equation}
f_{s\text{ }}=\frac SA,  \label{32}
\end{equation}
Greiner and his co-workers had predicted that there exists a critical $f_{sc%
\text{ }}$at zero temperature and pointed that the metastable or stable
strangelets could only be found above this value $f_{s\text{ }}>f_{sc\text{. 
}}$Our results strongly support their prediction and extend it to the finite
temperature. We obtained $f_{sc}=1$ at $T=50\mbox{MeV}$. This value does not
depend on the baryon numbers. But as shown in figure 14, $f_{sc}$ depends on
the temperature. It becomes $f_{sc}=%
{\displaystyle {10 \over 5}}%
=2$ at $T=20\mbox{MeV}$.

Now we turn to discuss the charge dependence of the stability of strangelet.
As shown in figures 12, 13 and 14, higher negative charge is in favour of
the metastable and/or stable strangelets. A maximal charge, which is $Z_m=1$
in figure 12 ($A=5,T=50\mbox{MeV}$), $Z_m=6$ in figure 13 ($A=10,T=50%
\mbox{MeV}$) and $Z_m=-3$ in figure 13 ($A=5,T=20\mbox{MeV}$), respectively,
exists. The strangelets would be stable or metastable when $Z\leq Z_m$.
Defining charge fraction $f_z$ as 
\begin{equation}
f_{z\text{ }}=\frac ZA,  \label{33}
\end{equation}
We find that the maximal charge fraction depends on not only the temperature
but also the baryon number $A$. For example, $f_{zm}$ equals to $%
{\displaystyle {1 \over 5}}%
$ for $A=5$ and $%
{\displaystyle {6 \over 10}}%
$ for $A=10$ when $T=50\mbox{MeV}$.

\subsection{Stable strangelet}

The strangelet which is stable against both strong and weak decay is called
stable strangelet. The stable strangelets are shown in figure 12, 13 and 14
by filled squares. We find that only a few strangelets can not decay via
strong and weak reactions and be completely stable. For example, as shown in
figure 12, for $A=5$ and $T=50\mbox{MeV}$, in total $128$ strangelets, only $%
4$ strangelets are stable ($Z=-3,S=12$), ($Z=-4,S=13$), ($Z=-4,S=14$) and ($%
Z=-5,S=15$).

Finally, we hope to emphasize that the high negative charge and the high
strangeness are in favour of the stable strangelets obviously at finite
temperature. The stable strangelets in figures 12, 13 and 14 are all highly
negative charged. The conclusion given by Greiner et. al at zero temperature 
\cite{8} is still correct at finite temperature.

\section{Summary}

In summary, by using the QMDTD\ model, we have studied the thermodynamical
properties and the stability of strangelets at finite temperature. We obtain:

(1), For fixed strangeness and charge, free energy per baryon decreases as
the temperature increases. The stable radius of the strangelet decreases
when the strangeness increases or the charge decreases.

(2), The higher temperature is in favour of the stability of the
strangelets. Comparing figure 12 with figure 14, we see clearly that the
total area of metastable and stable strangelets expands when the temperature
increases.

(3), The higher negative charge and higher strangeness number are in favour
of the stability of the strangelets. The stable strangelets are highly
negative charged and have high strangeness.

\section{Acknowledgment}

This work is supported in part by the NNSF of China under contract Nos.
19975010, 10047005 and 19947001.

\section{Figure Captions}

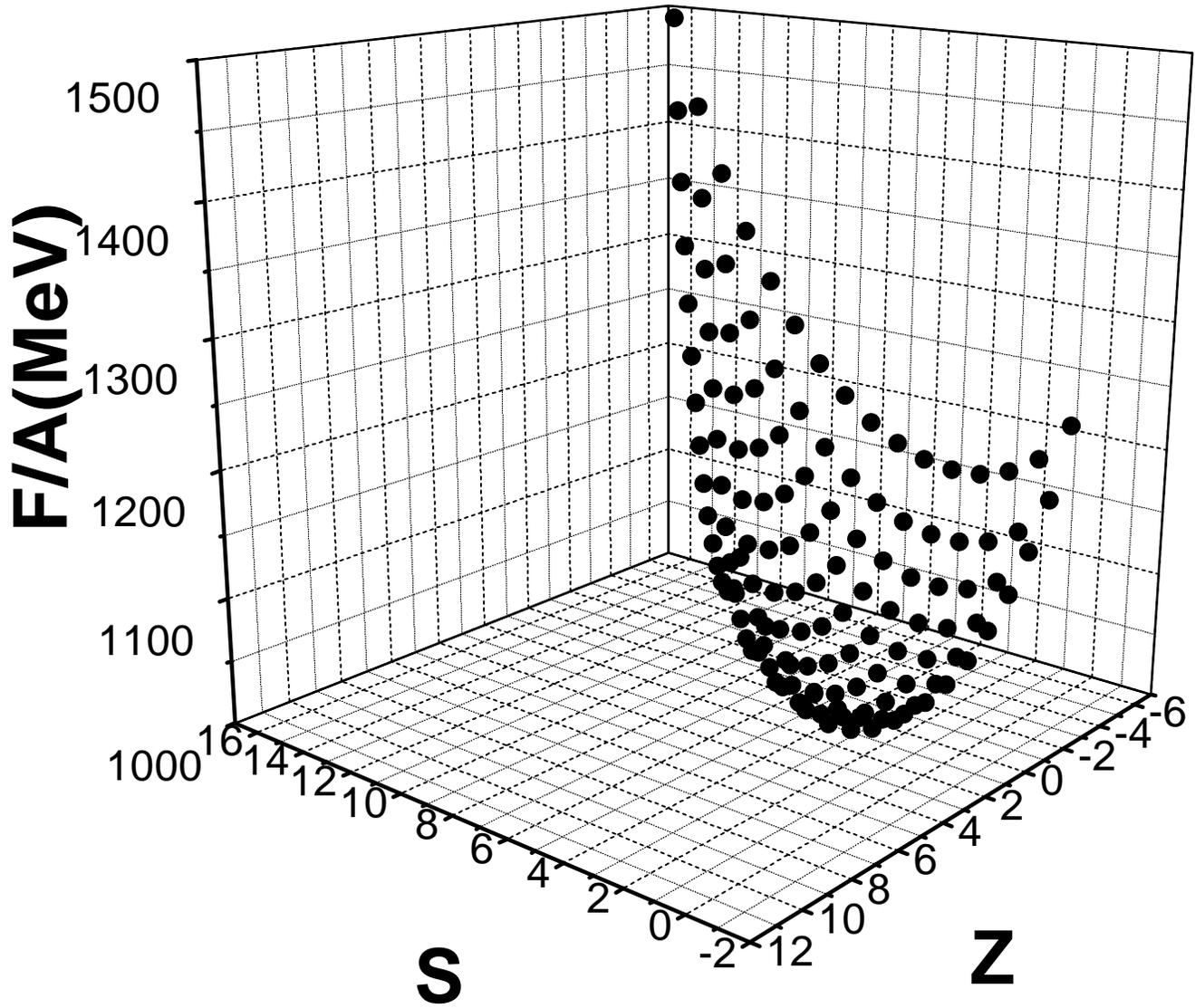
Figure 1. The dots stand for the free energy per baryon $F/A$ in a $S-Z$
plane for all possible strangelets with $A=5$ and $T=20\mbox{MeV}$.

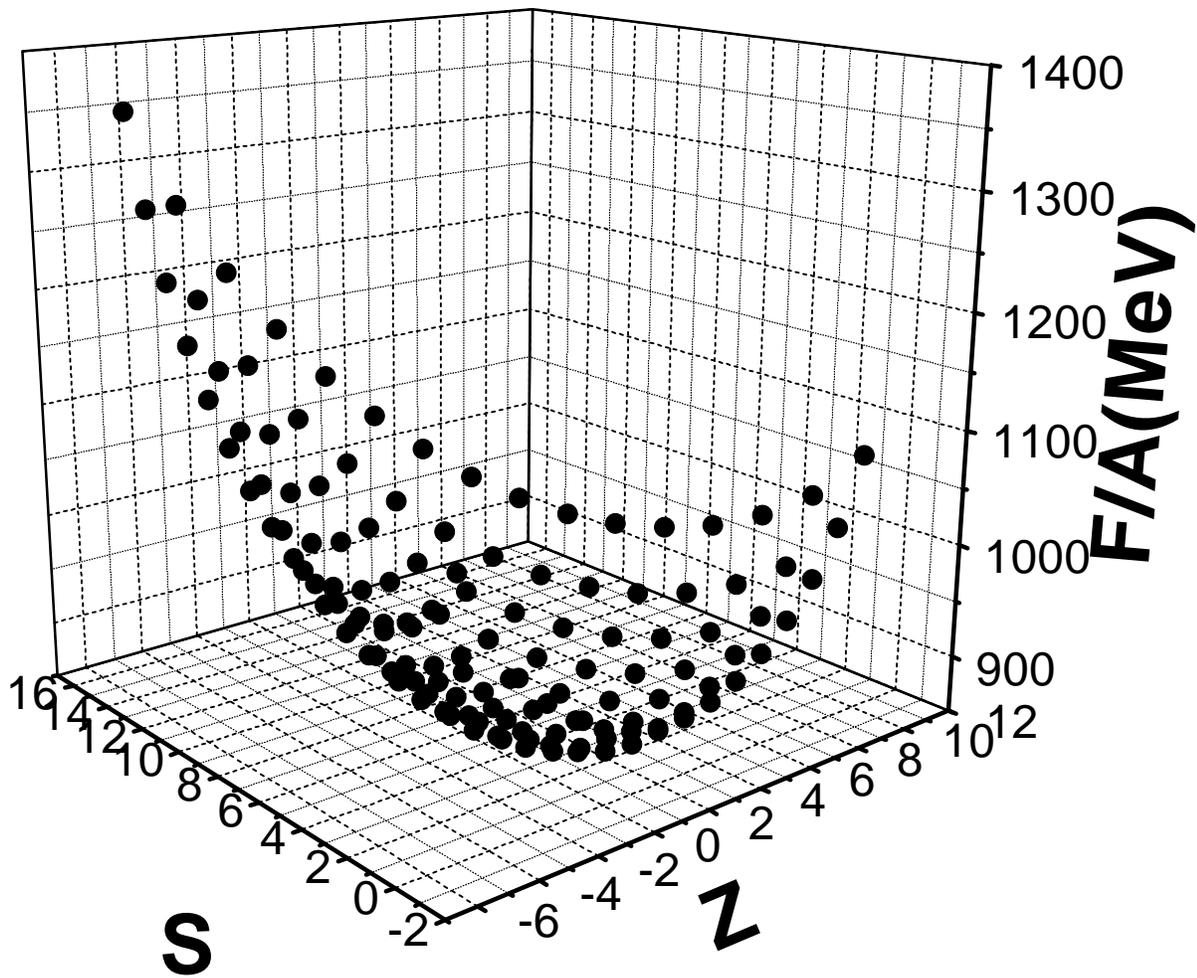
Figure 2. The same as figure 1, but for the temperature $T=50\mbox{MeV}$.

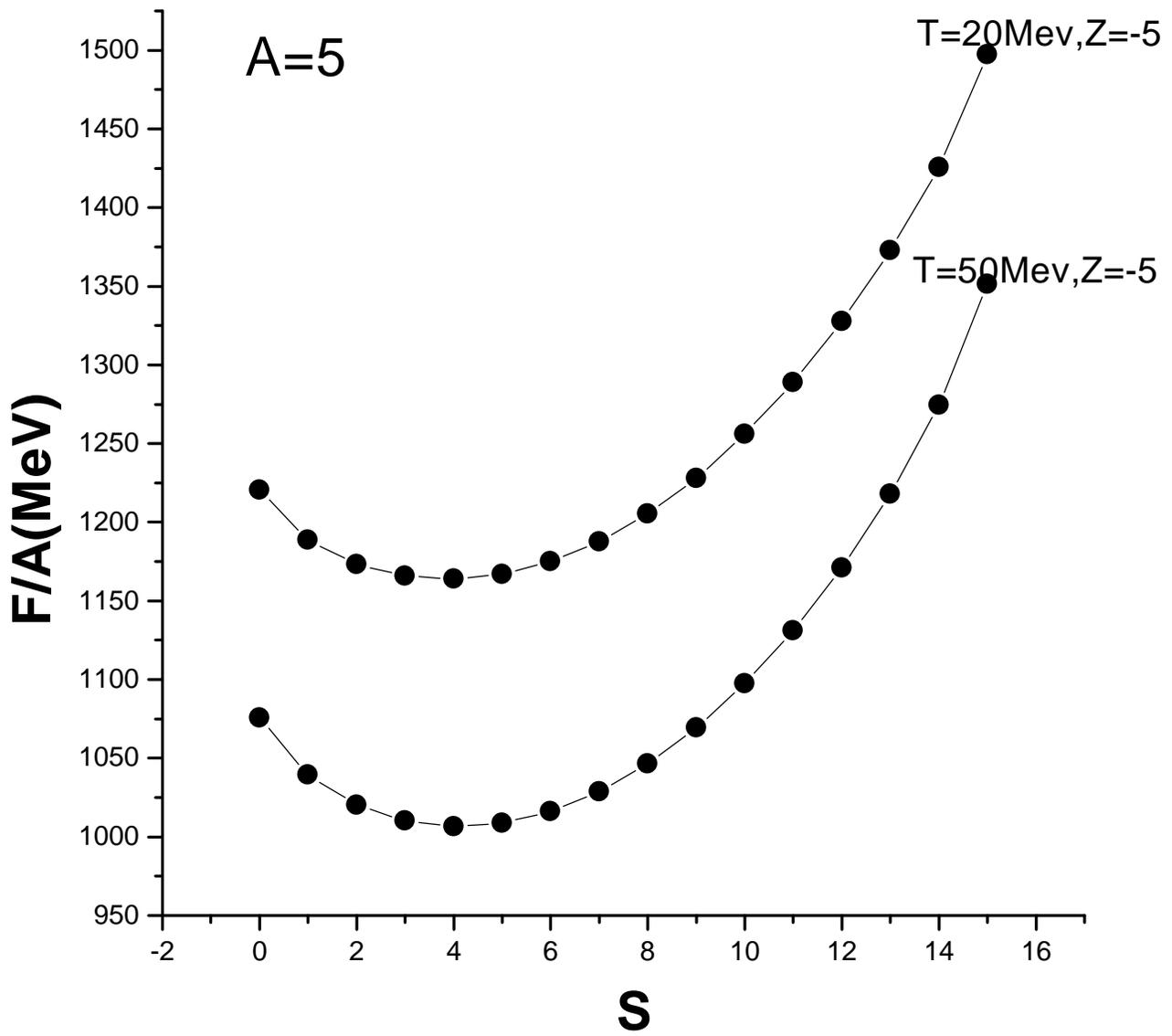
Figure 3. The free energy per baryon $F/A$ as functions of strangeness
number $S$ with $A=5$ and $Z=-5$, the two lines represent the temperature $%
T=20\mbox{MeV}$ and $50\mbox{MeV}$, respectively.

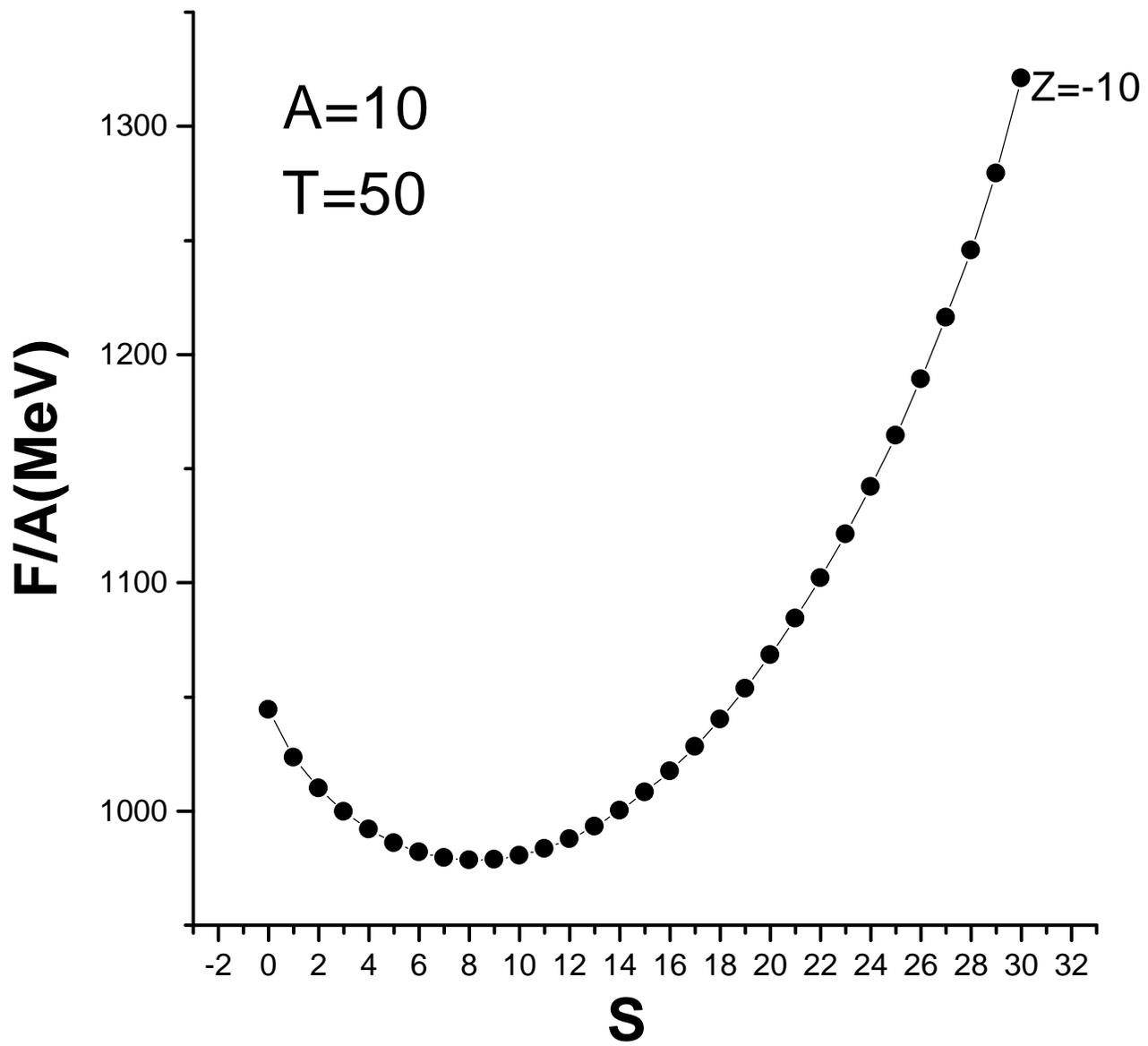
Figure 4. The free energy per baryon $F/A$ as a function of strangeness
number $S$ with $A=10$ , $Z=-10$ and $T=50\mbox{MeV}$.

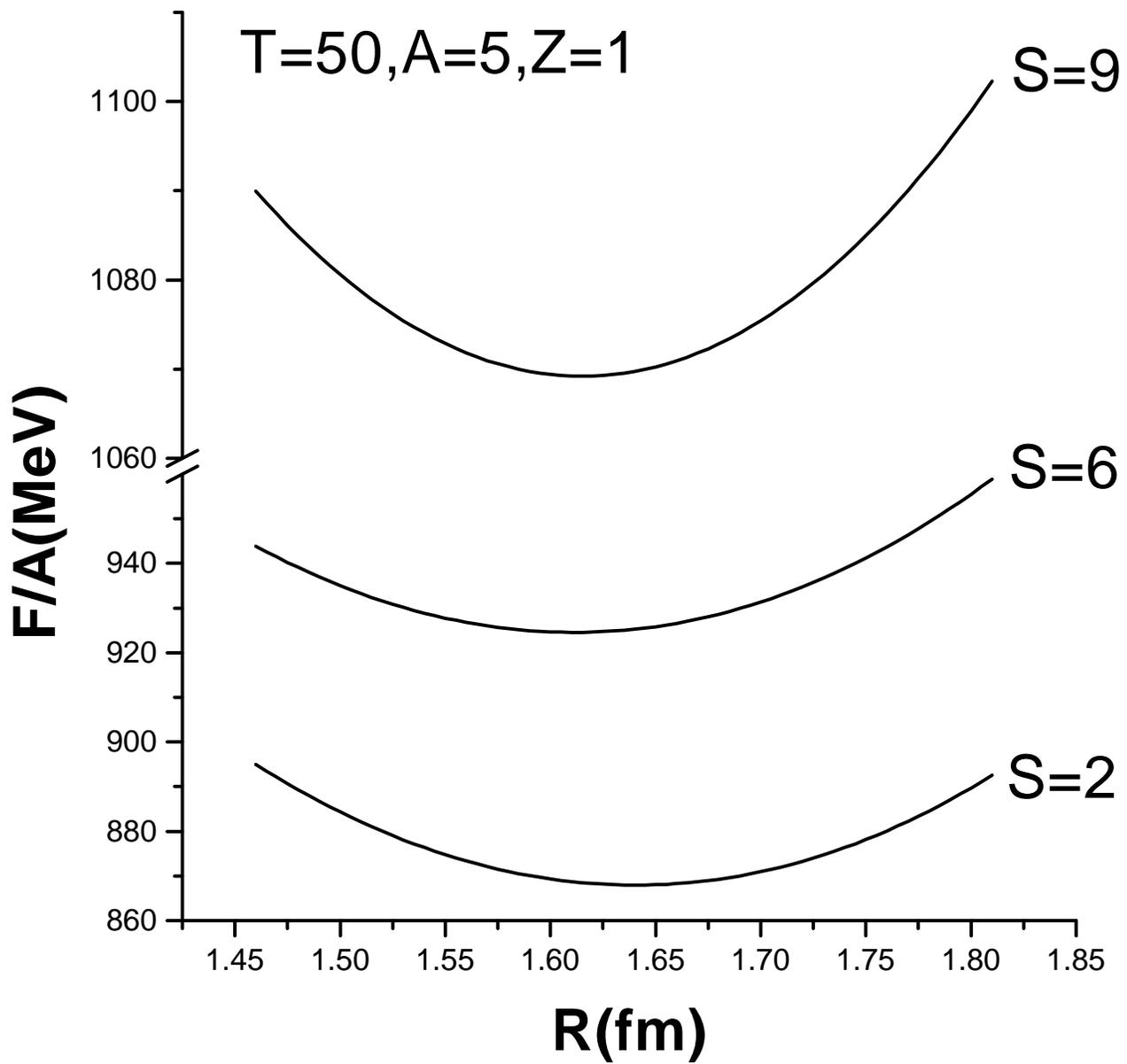
Figure 5. The free energy per baryon $F/A$ as functions of radius $R$ with $%
A=5$ , $Z=1$ and $T=50\mbox{MeV}$, the three lines represent the strangeness
number $S=2,6$ and $9$, respectively.

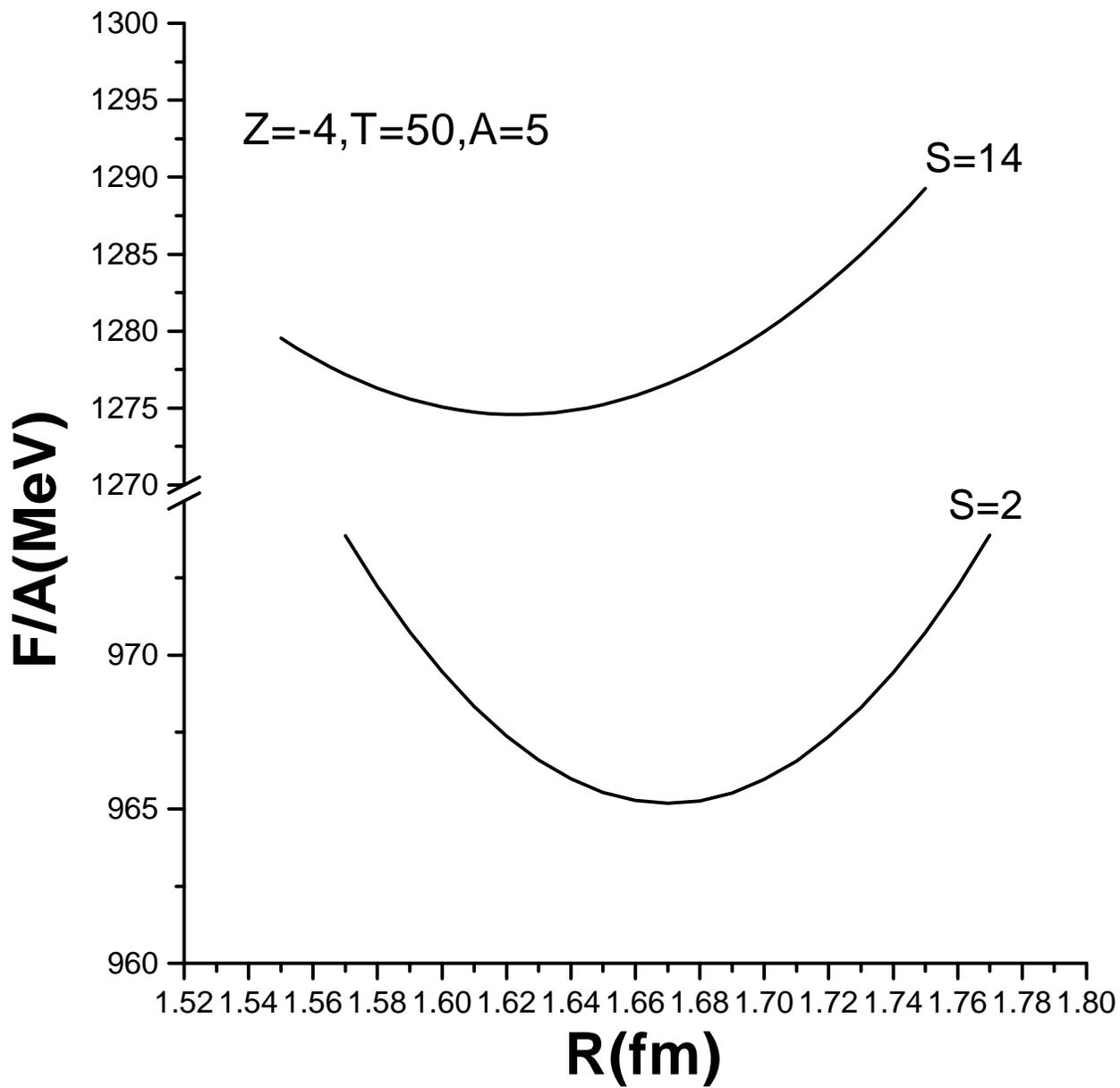
Figure 6. The free energy per baryon $F/A$ as functions of radius $R$ with $%
A=5$ , $Z=-4$ and $T=50\mbox{MeV}$, the two lines represent the strangeness
number $S=2$ and $14$, respectively.

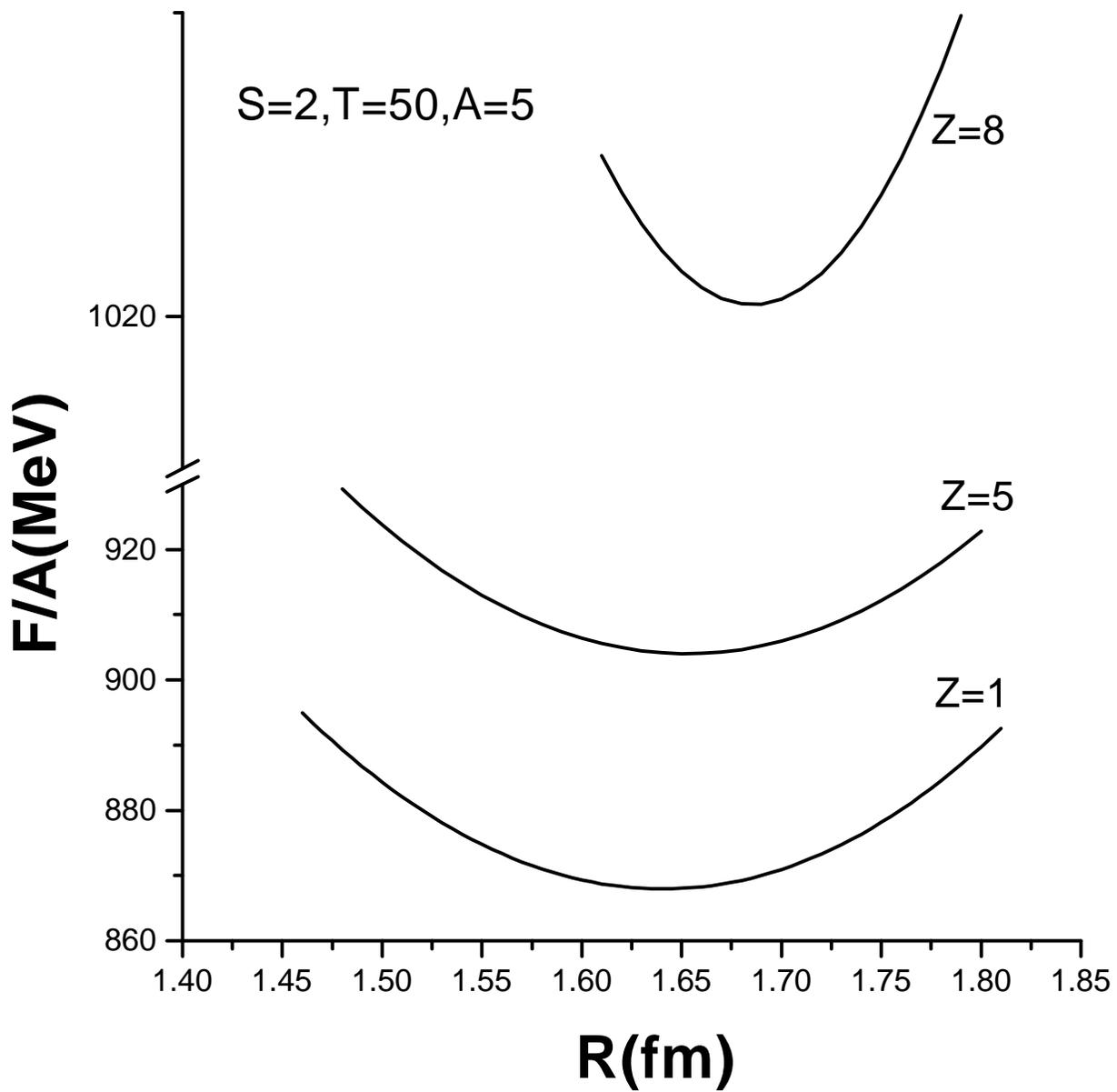
Figure 7. The free energy per baryon $F/A$ as functions of radius $R$ with $%
A=5$ , $S=2$ and $T=50\mbox{MeV}$, the three lines represent the electric
charge $Z=1,5$ and $8$, respectively.

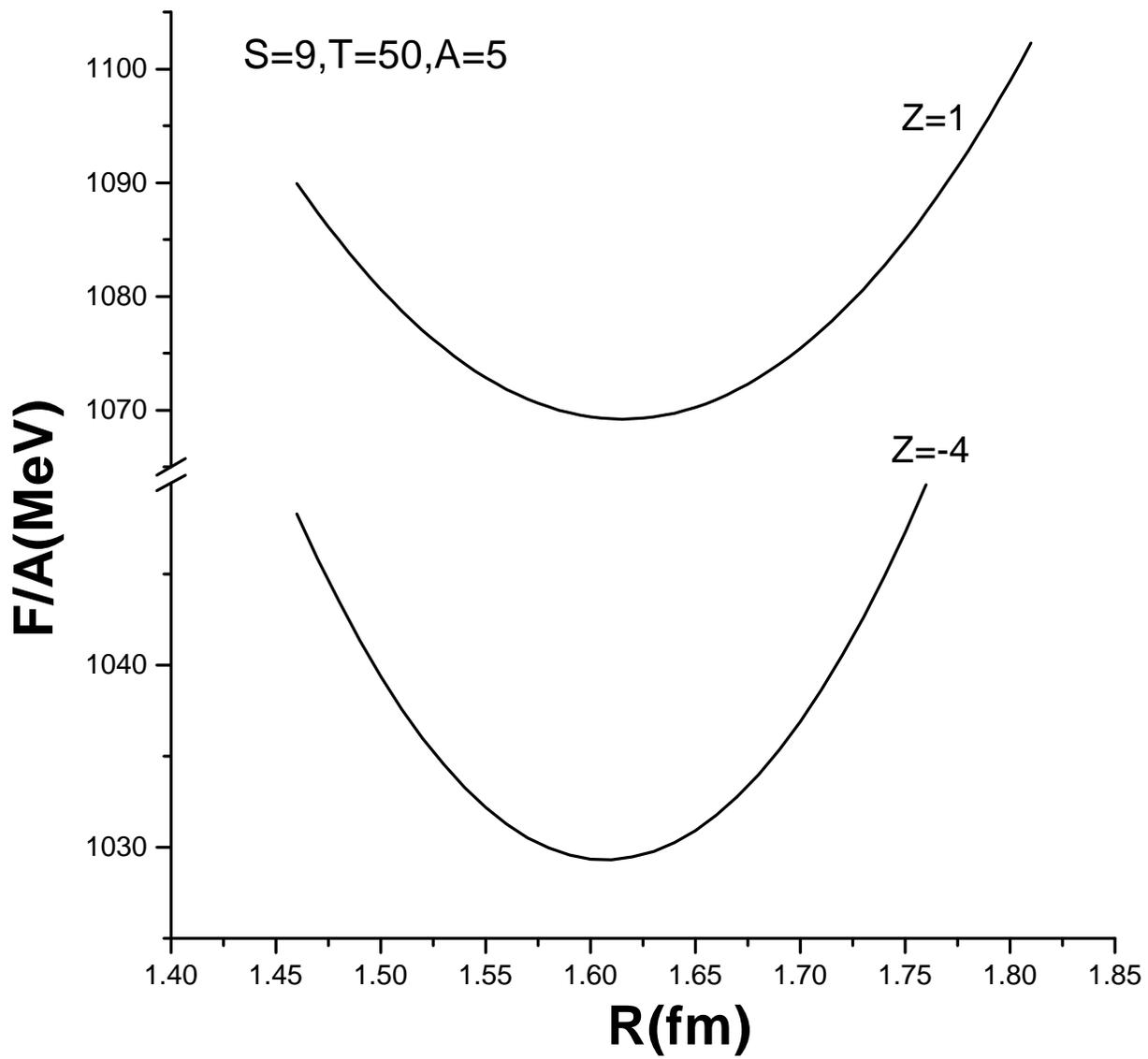
Figure 8. The free energy per baryon $F/A$ as functions of radius $R$ with $%
A=5$ , $S=9$ and $T=50\mbox{MeV}$, the two lines represent the electric
charge $Z=-4$ and $1$, respectively.

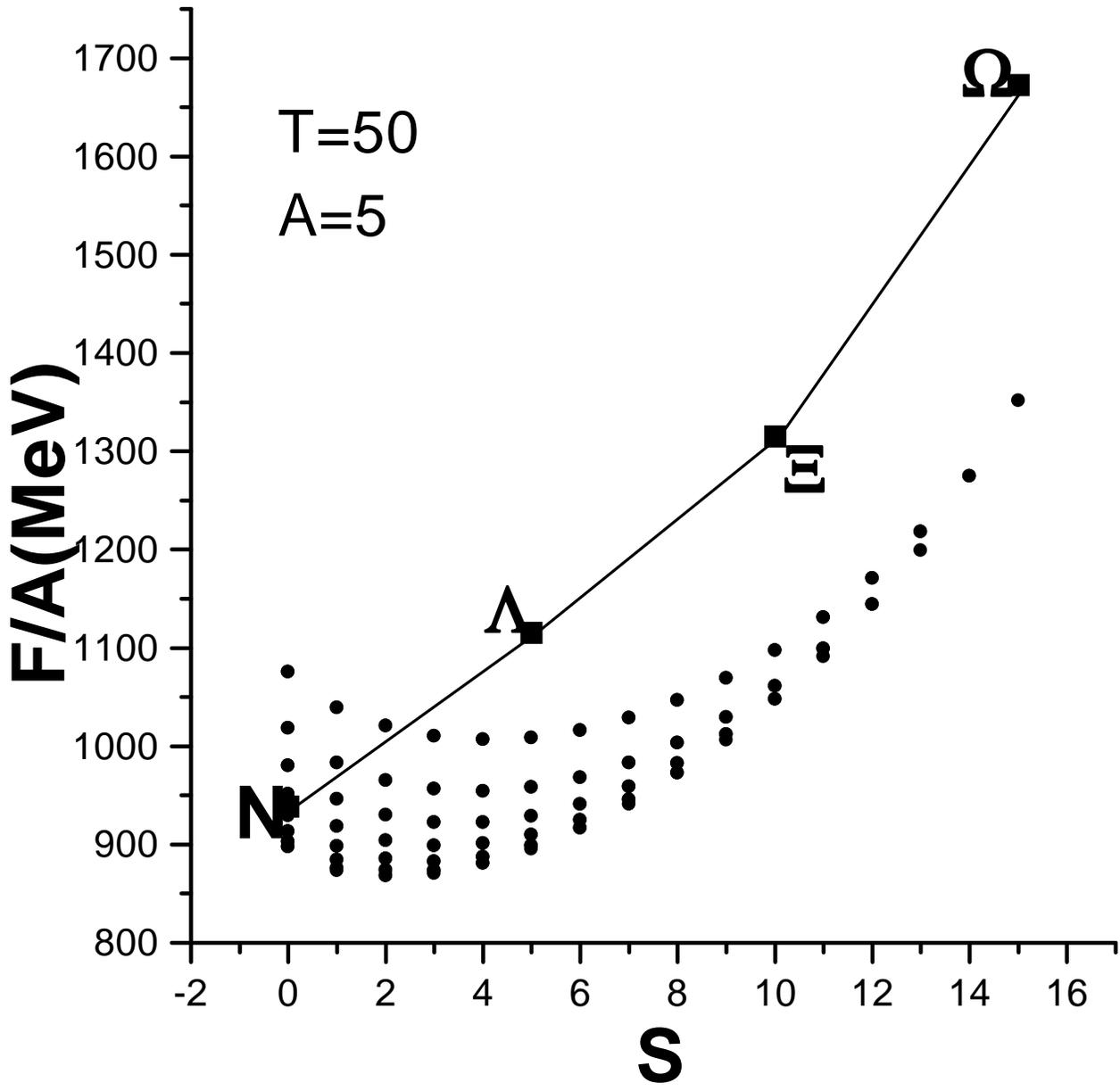
Figure 9. The dots stand for the free energy per baryon $F/A$ with various
strangeness numbers $S$ for all possible strangelets with baryon number $A=5$
at temperature $T=50\mbox{MeV}$. The masses of nucleon, $\Lambda ,\Xi $ and $%
\Omega $ are represented by the filled squares, respectively.

Figure 10. The same as figure 9, but for baryon number $A=10$.

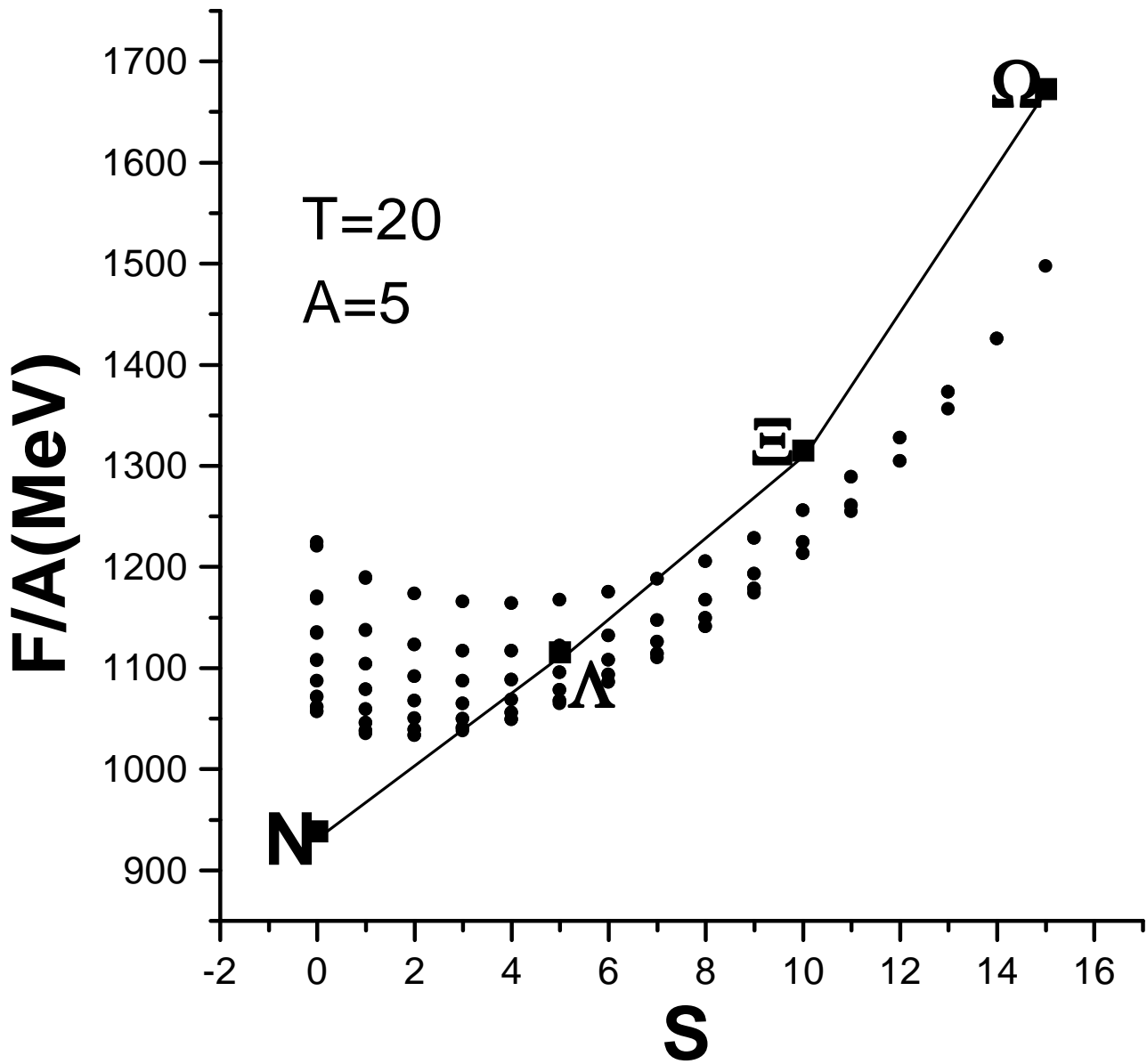
Figure 11. The same as figure 9, but for temperature $T=20\mbox{MeV}$.

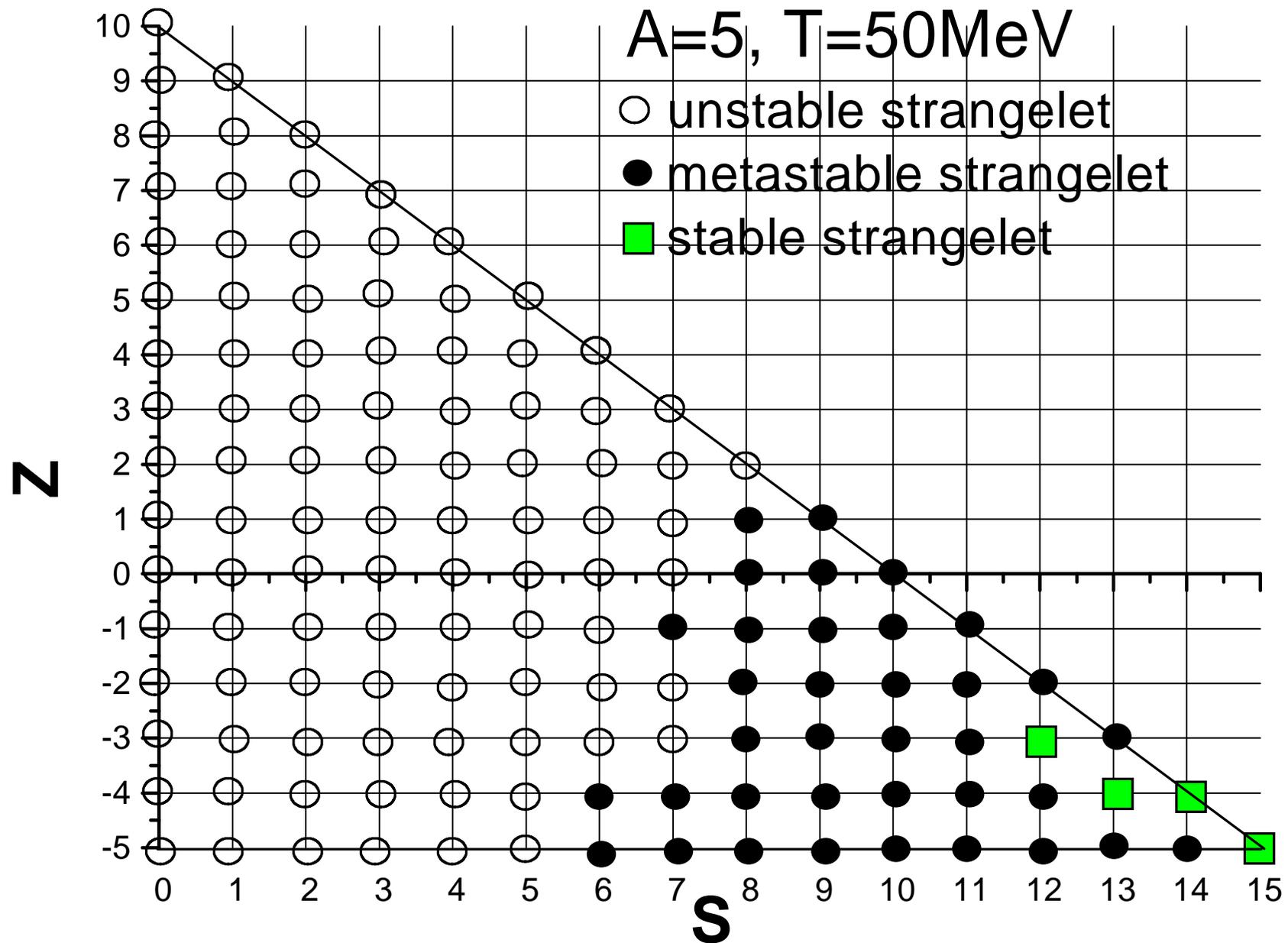
Figure 12. The electric charge $Z$ as a function of the strangeness number $%
S $ for unstable strangelets (open circles), metastable strangelets (filled
circles) and stable strangelets (filled squares) with baryon number $A=5$ at
temperature $T=50\mbox{MeV}$.

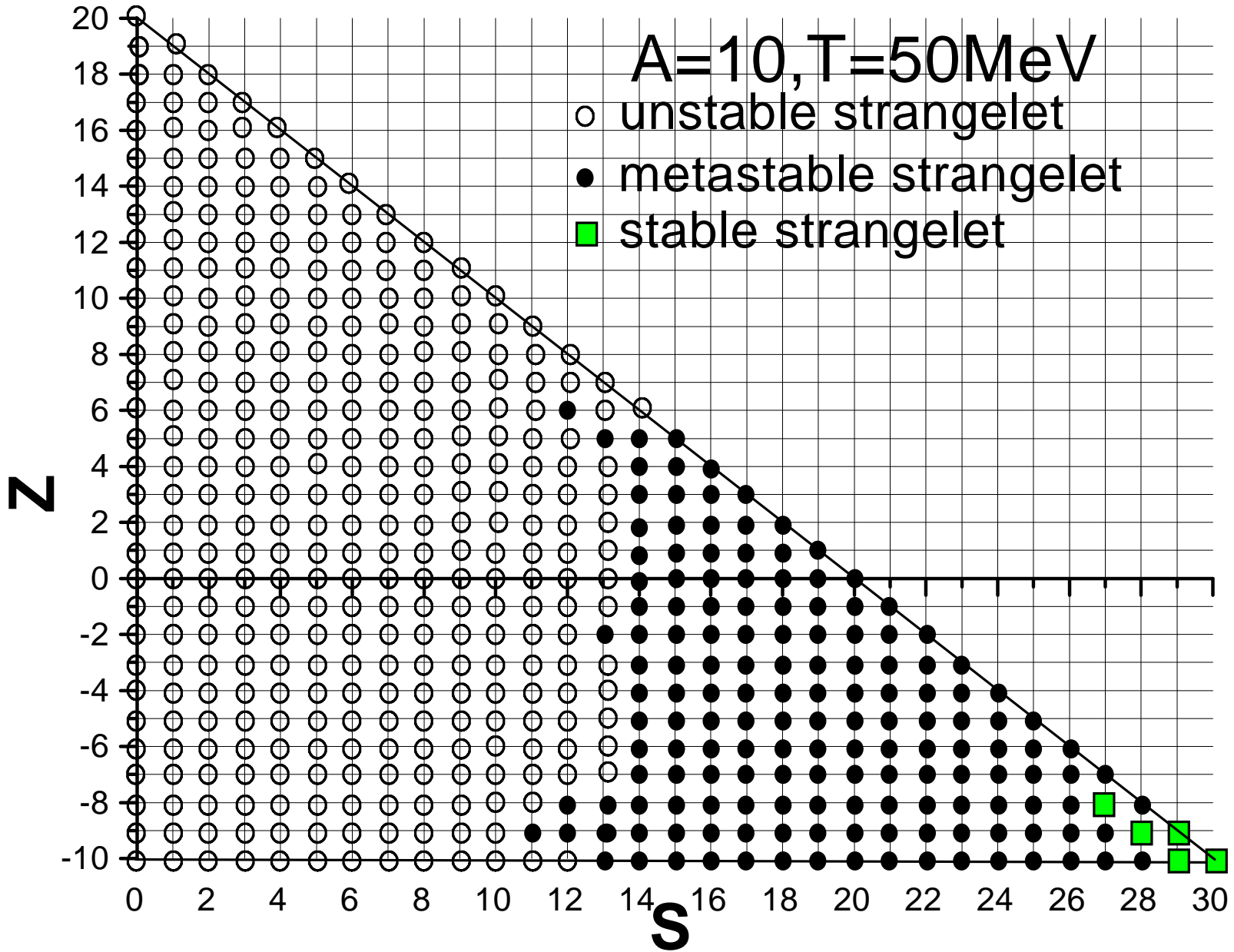
Figure 13. The same as figure 12, but for baryon number $A=10$.

Figure 14. The same as figure 12, but for temperature $T=20\mbox{MeV}$.

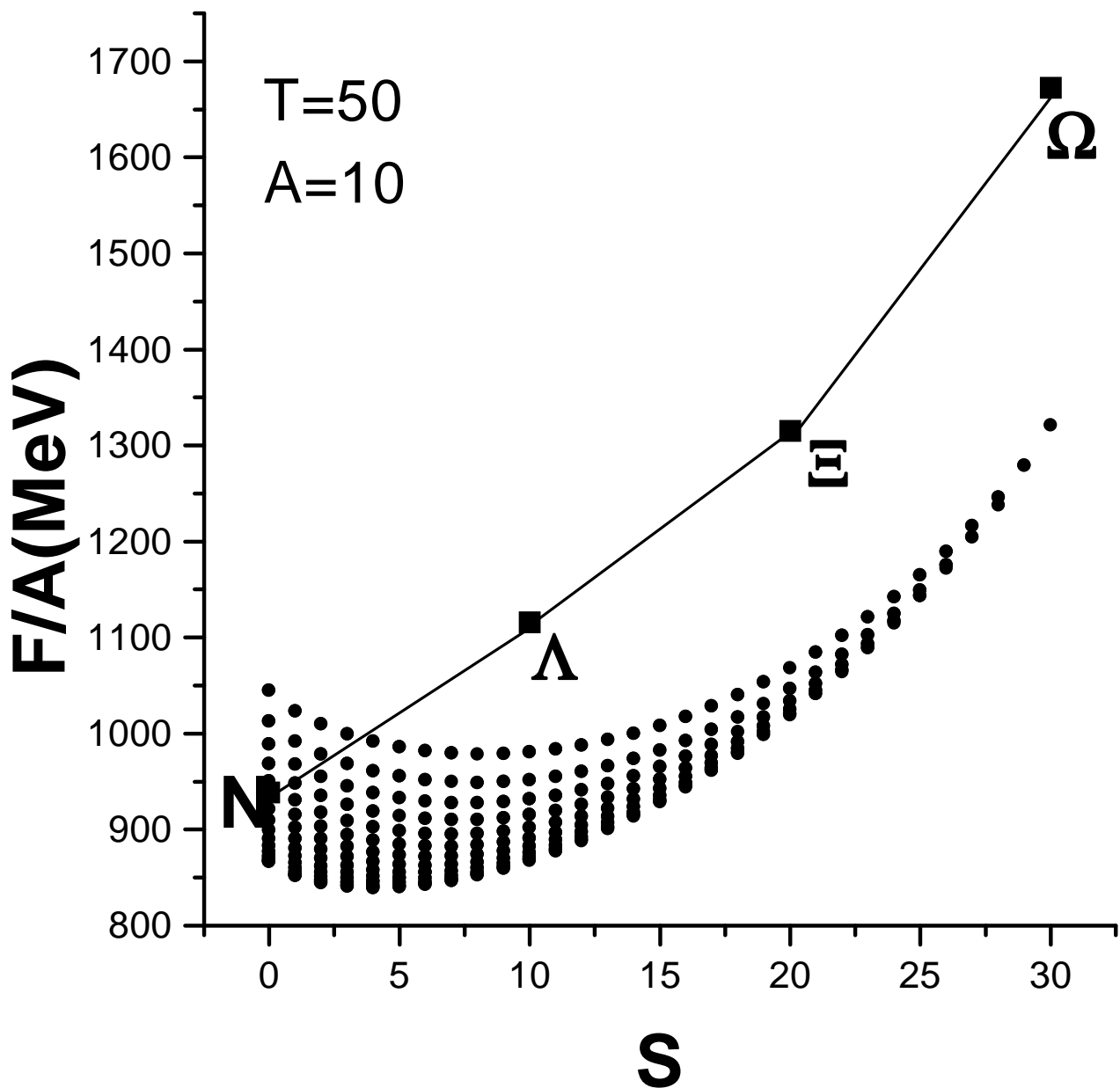
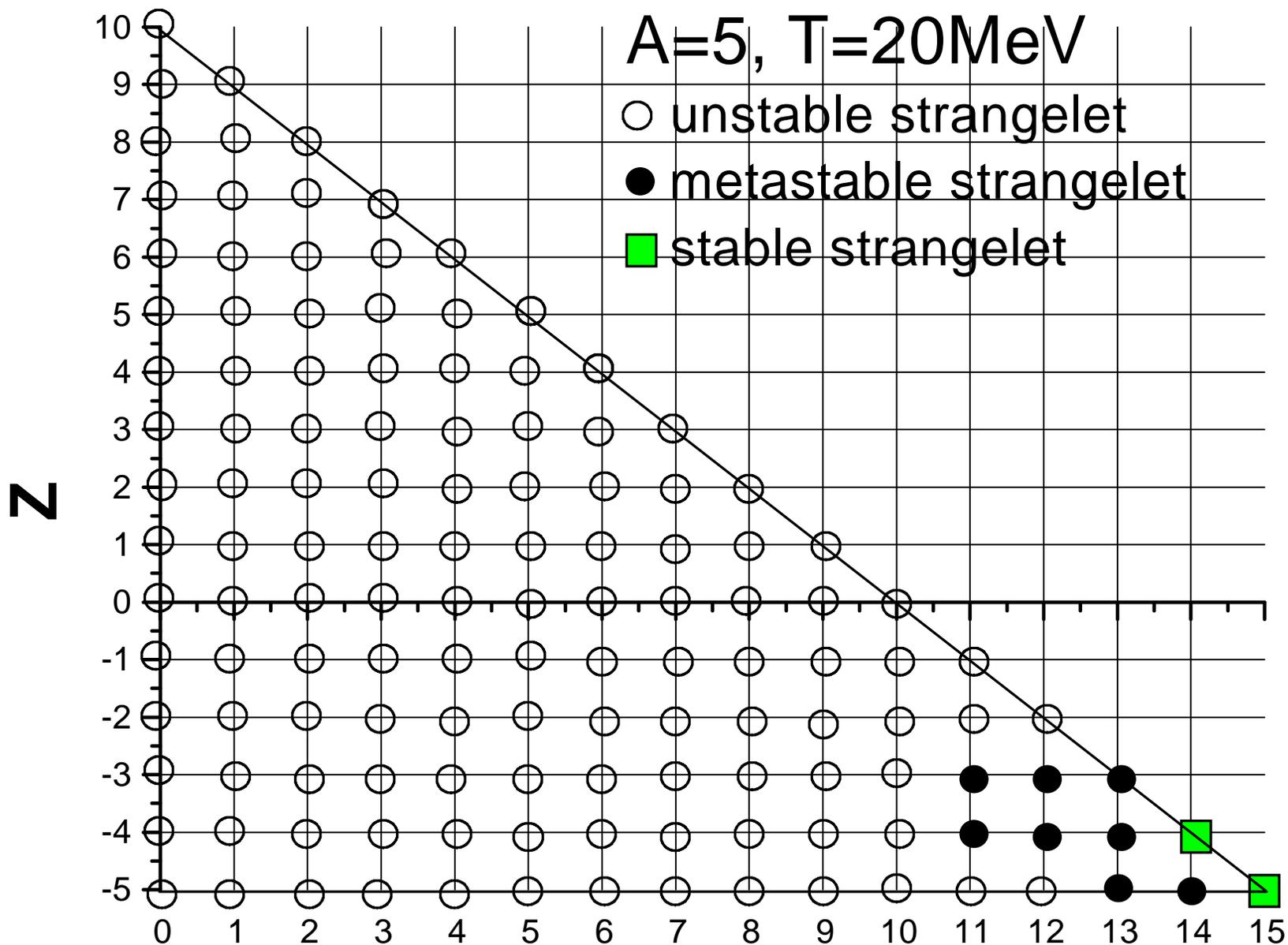
\end{document}